\documentclass[12pt]{iopart}
\usepackage{epsf}
\usepackage{graphicx}
\def\simlt{\lower.5ex\hbox{$\; \buildrel < \over \sim \;$}}

\def\mnras{{\it Month.\ Not.\ Roy.\ Astron.\ Soc.\ }}

\def\prd{{\it Phys.\ Rev.\ D}}

\def\be{\begin{eqnarray}}
\def\ee{\end{eqnarray}}
\def\bi{\bibitem}

\begin{document}

\title[A universal VDF]
{A universal velocity distribution of relaxed collisionless structures}

\author{Steen H. Hansen, Ben Moore, Marcel Zemp \& Joachim Stadel}
\address{University of Zurich, Winterthurerstrasse 190, 8057
Zurich, Switzerland} 

\begin{abstract}
Several general trends have been identified for
equilibrated, self-gravitating collisionless systems, such as density
or anisotropy profiles.  These are integrated quantities which
naturally depend on the underlying velocity distribution function
(VDF) of the system.  We study this VDF through a set of numerical
simulations, which allow us to extract both the radial and the
tangential VDF. We find that the shape of the VDF is universal, in the
sense that it depends only on two things namely the dispersion 
(radial or tangential) and the
local slope of the density. Both the radial and the tangential VDF's
are universal for a collection of simulations, including controlled
collisions with very different initial conditions, radial infall
simulation, and structures formed in cosmological simulations.
\end{abstract}




\maketitle

\section{Introduction\label{sec:intro}}

The last few years have shown remarkable progress in the understanding
of pure dark matter structures. This has been provided initially by
numerical simulations which have observed general trends in the
behaviour of the radial density profile of equilibrated dark matter
structures from cosmological simulations, which roughly follow an NFW
profile~\cite{nfw96,moore}, see \cite{diemand04} for references.
General trends in the radial dependence of the velocity anisotropy has
also been suggested~\cite{cole,carlberg}.  More recently have more
complex relations been identified, which even hold for systems which
do not follow the most simple radial behaviour in density. These
relations are first that the phase-space density, $\rho/\sigma^3$, is
a power-law in radius~\cite{taylor}, and second that there is a linear
relationship between the density slope and the
anisotropy~\cite{hm}.

All of these are integrated quantities, and still very little is known
about the underlying velocity distribution function (VDF) of real
collisionless systems. We will here initiate such a study by
performing a set of different simulations of purely collisionless
systems. We will then extract the VDF and split this into the radial
part and the tangential part for numerous bins in potential energy.
This will allow us to more fully appreciate the complicated structure
of the VDF. We will show that within the structures which have been
perturbed strongly and subsequently allowed to relax, both the radial
and the tangential VDF's have a universal shape. This shape is given
only as function of the velocity dispersion
and the local density slope.

\section{Head-on collisions}
\label{sec:head}
The first controlled simulation is the head-on collision between two
initially isotropic NFW structures.  For the construction of the
initial structures, we use the Eddington method as described 
in~\cite{stelios}, which imposes an exponential cut-off in the
very outer region.  We create an initial NFW structure with zero
anisotropy containing 1 million particles.  The parameters are chosen
corresponding to a total mass of $10^{12}$ solar masses (concentration
of 10), and half of the particles are within 160 kpc. The structure is
in equilibrium in the sense that when evolving such a structure in
isolation its global properties remain unchanged.

We now place two such structures very far apart with 2000 kpc between the
centres, which is well beyond the virial radius. Using a softening of
0.2 kpc, we now let these two structures collide head-on with an initial
relative velocity of 100 km/sec towards each.  After several crossings
the resulting blob relaxes into a prolate structure.  We run all
simulations until there is no further time variation in the radial
dependence of the anisotropy and density.  We check that there is no
(local) rotation.  We run this simulation for 150 Gyrs, which means
that a very large part of the resulting structure is fully
equilibrated.

We now take the resulting structure (now containing 2 million
particles, in a prolate shape) and collide 2 copies together, again
with their centres separated by 2000 kpc, and with initial relative
velocity of 100 km/sec. We now use softening of 1 kpc.  The resulting
structure is even more prolate when observed in density contours, and
we evolve this for another 150 Gyr.

All simulations were carried out using PKDGRAV, a multi-stepping,
parallel code~\cite{stadel01}, which uses spline kernel softening and
multi-stepping based on the local acceleration of particles.  The
simulations were performed on the zBox at Zurich University.

\begin{figure}[hbt]
\begin{center}
\includegraphics[width=0.45\textwidth]{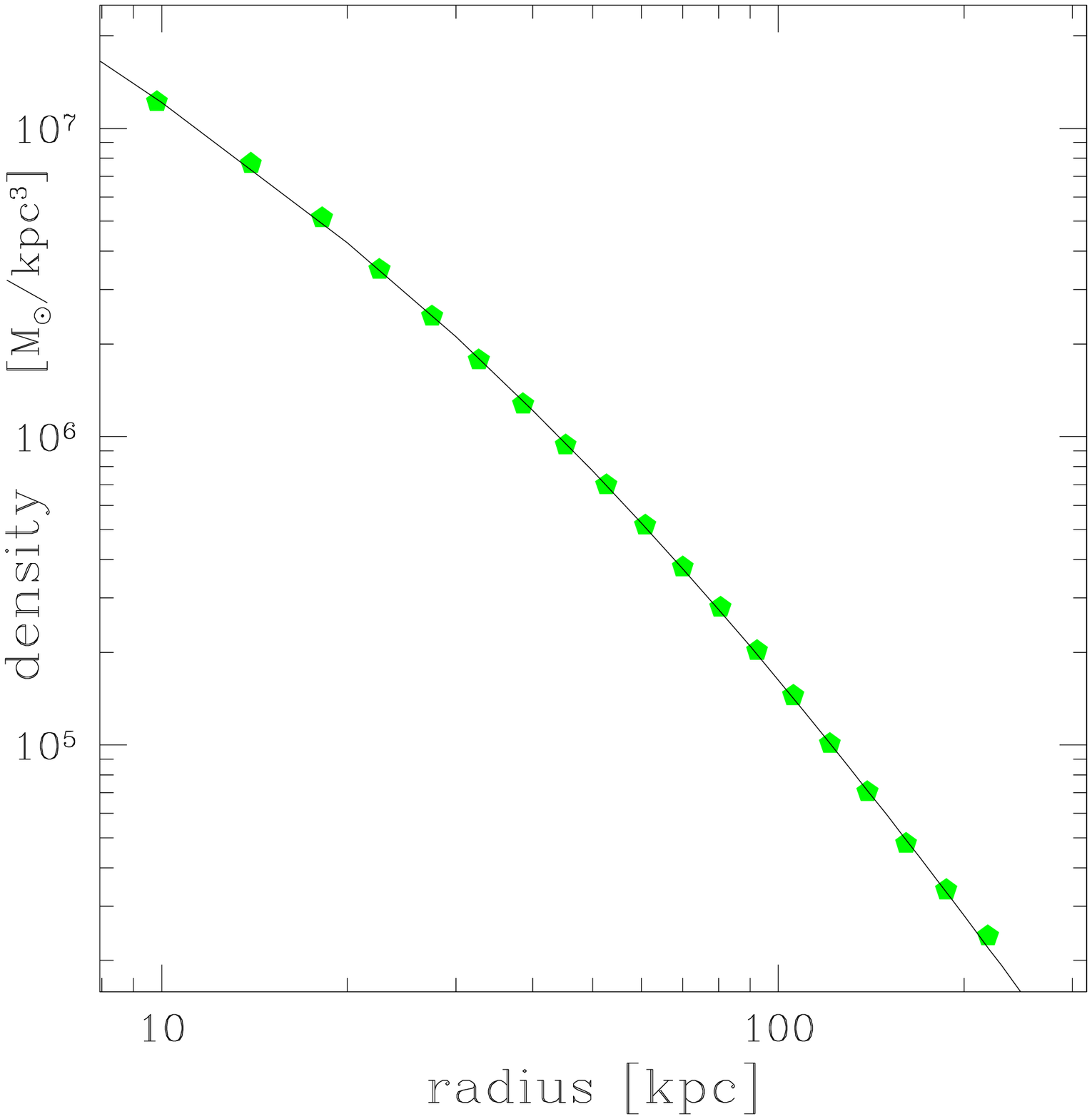}
\includegraphics[width=0.45\textwidth]{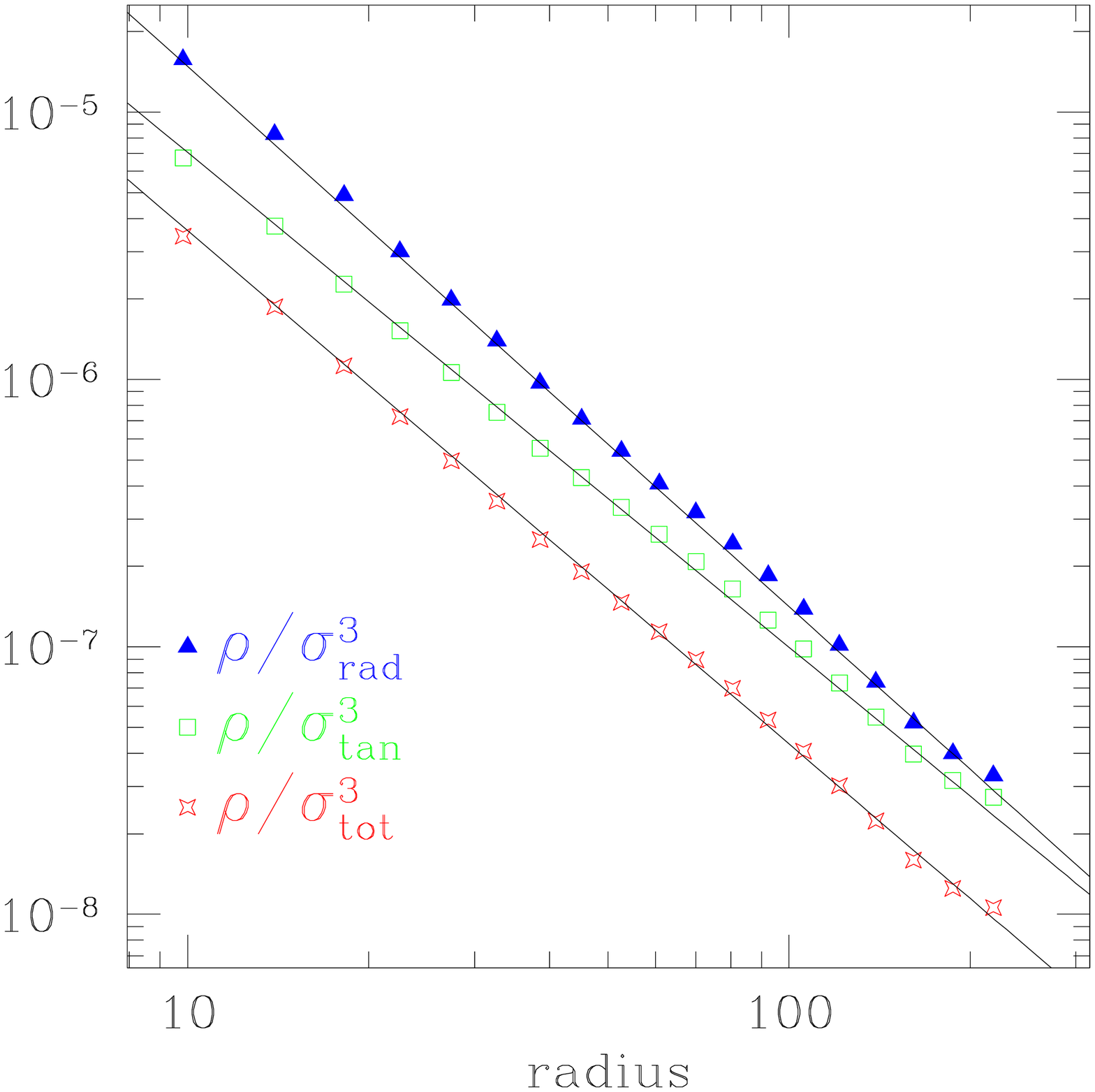}
\caption{Left panel: Resulting density profile of a 
repeated head-on collision between two initially isotropic NFW 
structures. The thin black line is of the NFW shape.
Right panel: Resulting phase-space density profile of a 
repeated head-on collision between two initially isotropic NFW 
structures. The thin black lines are just power-laws in radius,
with slopes $-2.02$ (radial), $-1.85$ (tangential), and
$-1.92$ (total).} \label{fig:dens}
\end{center}
\end{figure}

In order to analyze the resulting structure we divide it into bins
in potential energy, linearly distributed
from the softening length to beyond
the region which is fully equilibrated.
For the analysis and figures presented in this paper, we always only include
a region outside 3 times the softening. 
We center the structure on the center of potential
(using potential as weight-function instead of mass as done for center
of mass).
Most other analyses are centering on either the center of mass,
or on the most bound particle, but the difference is very small.
We calculate the local density for each particle from its nearest
32 neighbours, and we average the density for all the particles
in the potential bin. Similarly, each particle has a radius from
the center of potential, and we can average this over all the
particles in the potential bin. 
The resulting density
profile, $\langle \rho \rangle$ as a function of
$\langle r \rangle$,  is very similar to an NFW profile
(see figure~\ref{fig:dens}).

One can also calculate velocity dispersions for each potential bin of
the structure. We split the velocities into radial and tangential, and
on figure~\ref{fig:dens} we present the phase-space density,
$\rho/\sigma^3$, as a function of radius.  The $\sigma$ can be either
the radial, tangential or total.  The relationship between phase-space
density and radius has been considered several
times~\cite{taylor,dm,rasia,ascasibar,austin} and seems to be well
established.  Using the Jeans equation together with the fact that
phase-space density is a power-law in radius allows one to find the
density slope in the central region numerically~\cite{taylor}, and
even analytically for power-law densities~\cite{jeanspaper}. Recently
a much refined study \cite{dm} using both the phase-space density
being a power-law in radius and also the linear relationship between
density slope and anisotropy, showed that one can solve the Jeans
equations analytically and extract the radial dependence of density,
anisotropy, mass etc. This analysis was made using $\sigma_{\rm rad}$.
Whereas we here confirm that the phase-space density
roughly follows a power-law in radius, then it is not clear which of
the dispersions leads to a better fit to a power-law (see
figure~\ref{fig:dens}). Given the fact that the anisotropy
changes as function of radius,  
then if one of the phase-space densitities (e.g. using
$\sigma_{rad}$) is a power-law, then the other (e.g. using $\sigma_{tot}$)
must have slightly s-shaped residuals.

\begin{figure}[hbt]
\begin{center}
\includegraphics[width=0.32\textwidth]{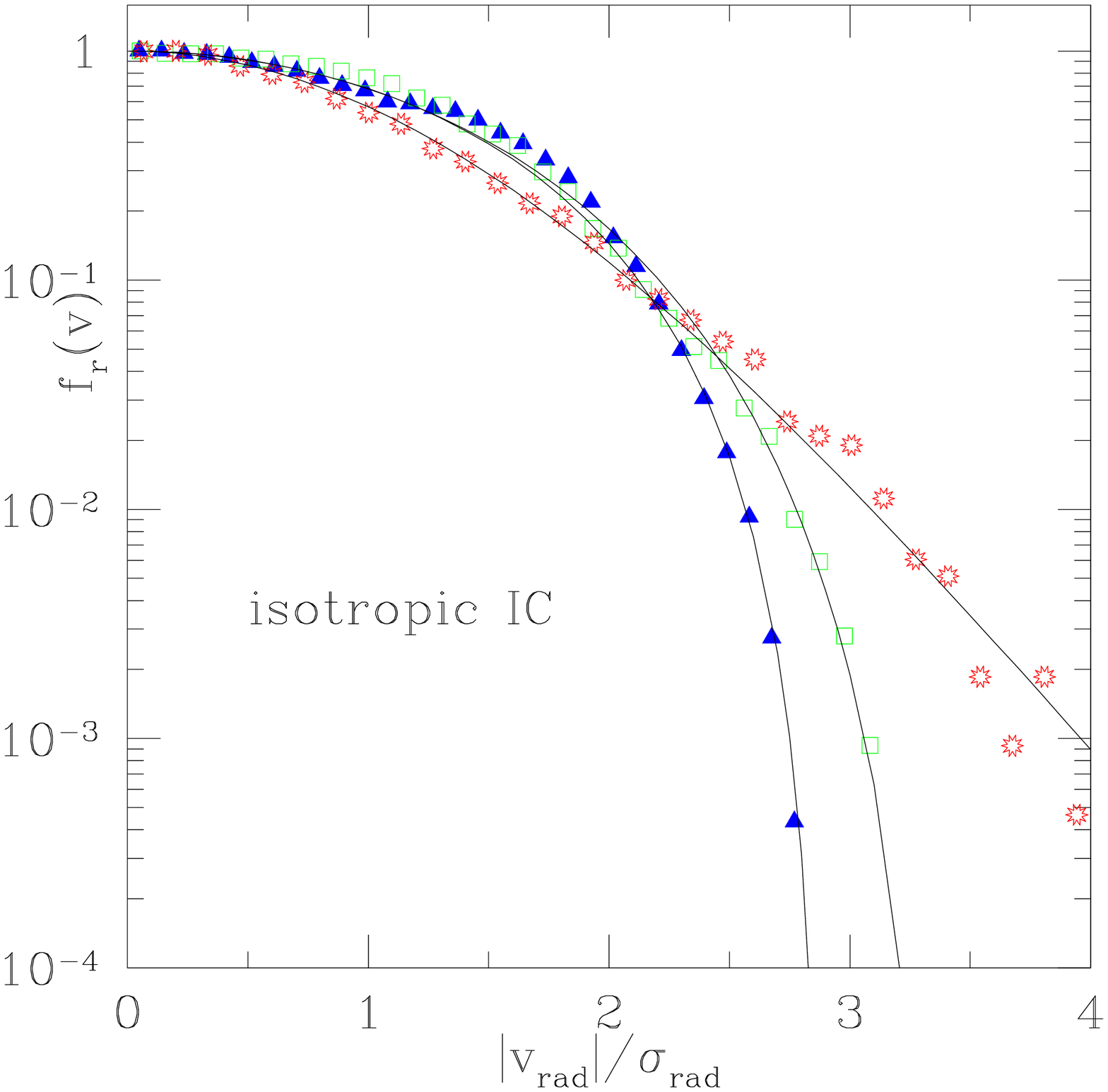}
\includegraphics[width=0.32\textwidth]{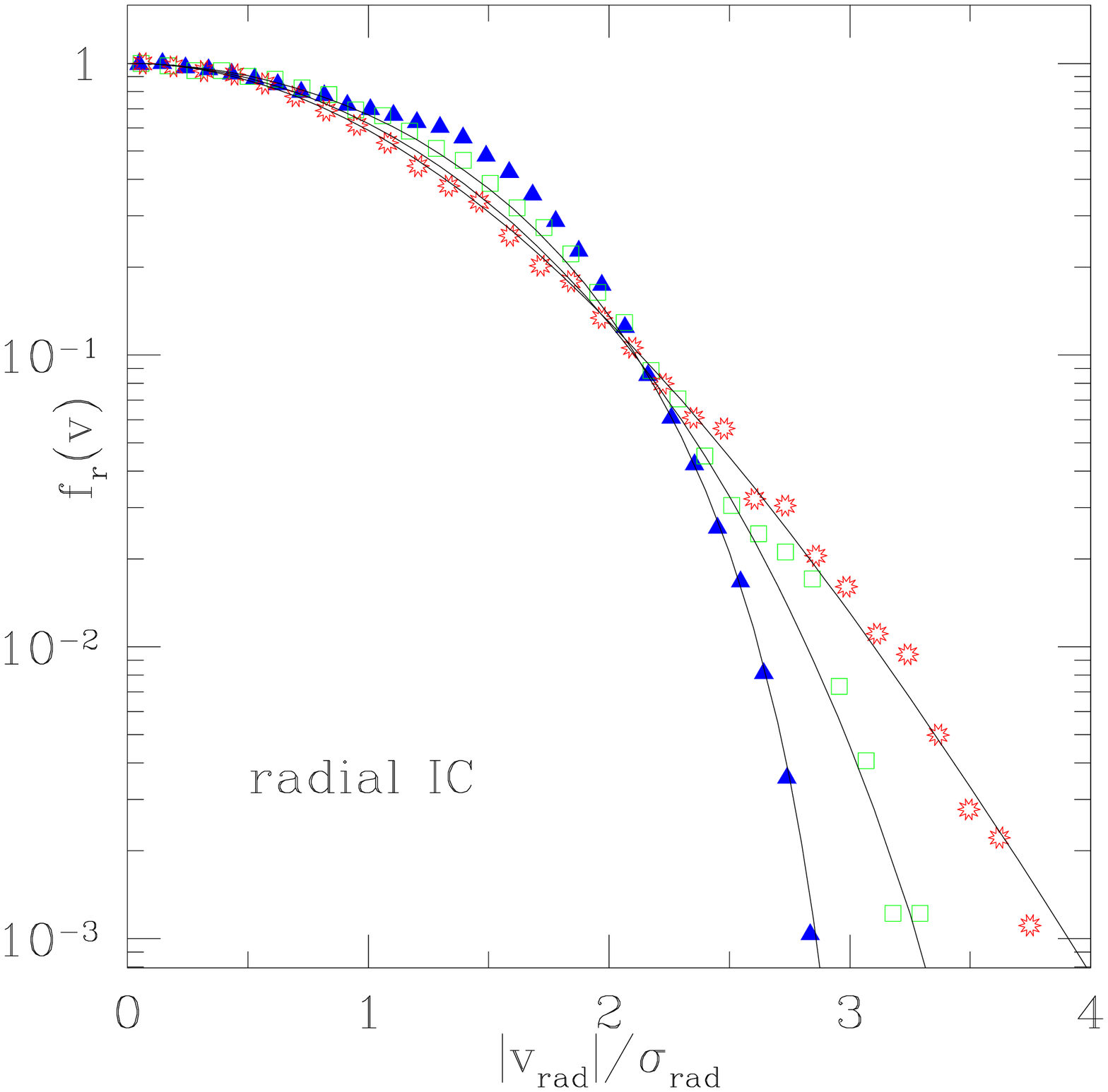}
\includegraphics[width=0.32\textwidth]{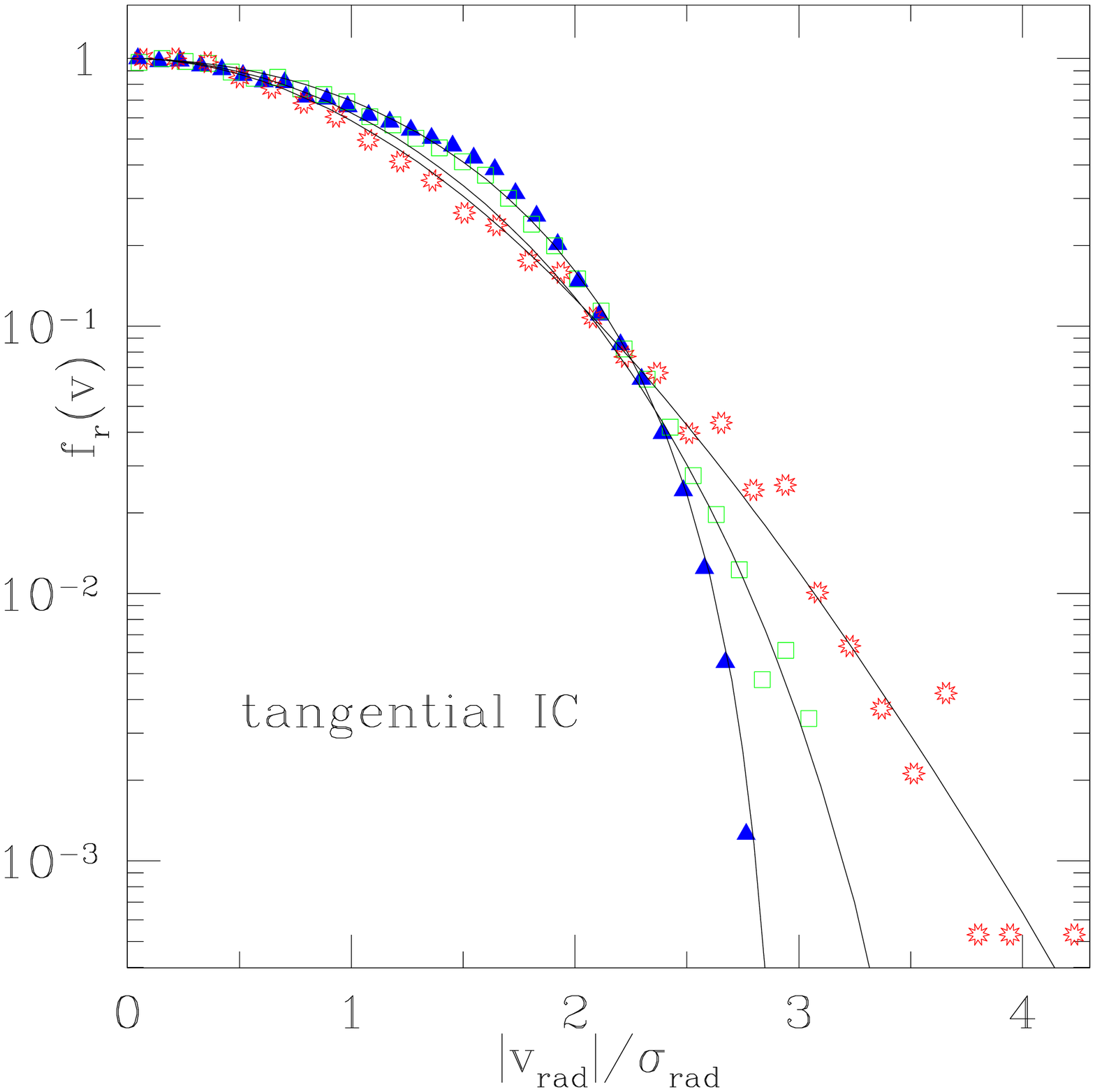}
\caption{The radial VDF for 3 potential bins in the simulation
of repeated headon collisions between NFW
structures with very different initial conditions. 
The potential bins are chosen near density slope of
$\alpha \approx -1.1$ (open red stars), 
$\alpha \approx -2$ (open green squares)
and $\alpha \approx -2.6$ (filled blue triangles).
The thin (solid) lines are of the shape given by 
eq.~(\ref{eq:tsallis}), using $q$'s of $0.75, 0.81$ and $1.08$ (fig a);
$q=0.79, 0.91$
and $1.05$ (fig b); and
$q=0.75, 0.89$
and $1.045$. The similarity between these figures is striking.}
\label{fig:rvdf}
\end{center}
\end{figure}

For each bin we can also define
the radial
derivative of the density (the density slope)
\begin{equation}
\alpha \equiv \frac{d {\rm ln} \langle \rho \rangle}{d {\rm ln} \langle
r \rangle} \, ,
\end{equation}
and we can extract the radial velocity distribution function (VDF) for
each potential bin. We present the radial VDF for bins with $\alpha
\approx -2.6, -2$ and $-1.1$ in the left panel of
figure~\ref{fig:rvdf}.  The red (open)
stars are from a bin in the region with shallow density slope ($\alpha
\approx -1.1$), the green (open) squares are from an intermediate bin
(slope close to $-2$), and the blue (filled) triangles are from a bin
in the outer region (slope close to $-2.6$). There seems to be a clear
trend, namely that the VDF from bins in the inner region have longer
high energy tails than VDF's from bins in the outer region. On the
same figure have we overplotted 3 black (solid) lines, which have the
shape
\be f_r(v) = \left( 1 - (1-q ) 
\, \left( \frac{v}{\kappa_1 \, \sigma_{\rm rad}} \right)^2 
\right) ^{\frac{q}{1-q}} \, ,
\label{eq:tsallis}
\ee 
where $q$ and $\kappa_1$ are free parameters.  This is of the Tsallis
form~\cite{tsallis88}, and depends on the entropic index $q$.  This
form reduces to the Gaussian shape $f(v) \sim {\rm exp}(-(v/v_0)^2)$
when $q=1$. This functional form is chosen partly for simplicity (and
because it includes the classical exponential), and partly because
some simple structures are know to have VDF's of exactly this
form~\cite{plastino,hansen05}. The corresponding $q$'s of these 3
lines in figure~\ref{fig:rvdf} 
are $0.75, 0.81, 1.08$, showing that the bin in the region with
more shallow density slopes (with $q=1.08$) has a longer high energy
tail than a Gaussian. In particular do we see that  
the resulting VDF's do not
have the shape of Gaussians with a cut-off at the escape velocity.

The shape of eq.~(\ref{eq:tsallis}) provides a reasonable fit to all
the energy range only for the most shallow profile, namely in the
spatial region where the velocity dispersion is isotropic. In the
regions with steeper density profile does the shape of
eq.~(\ref{eq:tsallis}) only provide a fit in the high energy region of
the VDF, and in order to fully fit the radial VDF in the low energy
region must one consider a more general shape. 
Detailed look at the figures in 
fig.~\ref{fig:rvdf} reveals that the general radial
VDF may be better fit with a more flat-topped (larger $q$) in the low
energy region, and with a (small) $q$ as described above for the high
energy region, however
we will not pursue this
issue further here.

It is worth commenting briefly on the definition of radial and
tangential used in this paper. We are splitting the equilibrated
structure in potential bins, which is because the energy of each
particle is composed of potential and kinetic energy. Since we wish to
extract the VDF's, then we must consider potential bins and not radial
bins, however, the potential bins are virtually spherical in the outer
region, and there is only a small effect in the central region. Now,
we define {\em "radial"} strictly with respect to a spherical
symmetry. Clearly the splitting between radial and tangential will
depend somewhat on the actual shape of the structure considered: if
the structure is very elongated, then the concept of {\em "radial"} must
be with respect to the shape of this structure. One can extract the
shape from either density contours or from the potential energy
contours, and the shape will be different (density contours are much
more triaxial), and hence the definition of {\em "radial"} depends on
which kind of shape one is referring to.  We will in this paper, for
simplicity, ignore this complication, and simply extract the {\em
radial} with respect to spherical symmetry. We hope to discuss this
issue in full in a future analysis.

\subsection{Dependence on initial conditions?}

To address the question of how strongly the VDF's  from the
head-on collisions described above depend on the initial conditions,
we now consider 2 simulations each with two steps.  i) First we create
a spherical isotropic NFW structure with 1 million particles as
described above.  We take each individual particle in this structure
and put its total velocity along the radial direction, maintaining the
sign with respect to inwards or outwards from the halo centre.  We
keep the energy of each particle fixed. This structure is thus
strongly radially anisotropic.  ii) Now we take two such radially
anisotropic structures and place their centres 2000 kpc apart (well
outside the virial radius) with relative velocity of 100 km/sec
towards each other.  We use softening of 1 kpc for these tests.
Before the centres of the two structures collide the radial motion of
the particles behaves almost like a radial infall simulation, and
strong scattering of the particles is observed.  The 2 individual
structures pick random orientations in space as they become
triaxial~\cite{merritt85}; their orientations are completely erased
after the two structures collide head-on. We let the structure
equilibrate. We take two copies of the resulting
structure and collide these head-on, again with 100 km/sec. The final
equilibrated region contains approximately 2.5 million particles (of
the total of 4 million particles).

The resulting VDF's extracted in potential bins near density slopes 
of $-2.3, -2, -1.2$ are shown  in the central panel of
figure~\ref{fig:rvdf}, and the
black (solid) lines are again of the shape of eq.~(\ref{eq:tsallis}),
with $q$'s of $0.79, 0.91$ and $1.05$.

Second, we run a simulation identical to the one just described,
except that we place each individual particle in the initial
structures on tangential orbits, without changing the energy of the
individual particles. This initial condition thus corresponds to
strongly tangential motion.  When letting this structure equilibrate
in isolation it settles down with a tangential anisotropy of $\beta
\sim -2$.  Again, two such structures are collided head-on. We again
take two copies of the resulting structure and collide these headon.
The resulting VDF's are shown in the right panel of 
figure~\ref{fig:rvdf}, where
the bins were chosen to be near density slope of $-2.6, -2$ and $-1.15$,
and the black (solid) lines are of the shape eq.~(\ref{eq:tsallis}),
with $q$'s of $0.75, 0.89$ and $1.045$.
One should keep in mind, that this last structure was created with 
zero radial velocity, and hence the radial VDF was created while
perturbing the structure and letting it relax.

\begin{figure}[hbt]
\begin{center}
\includegraphics[width=0.8\textwidth]{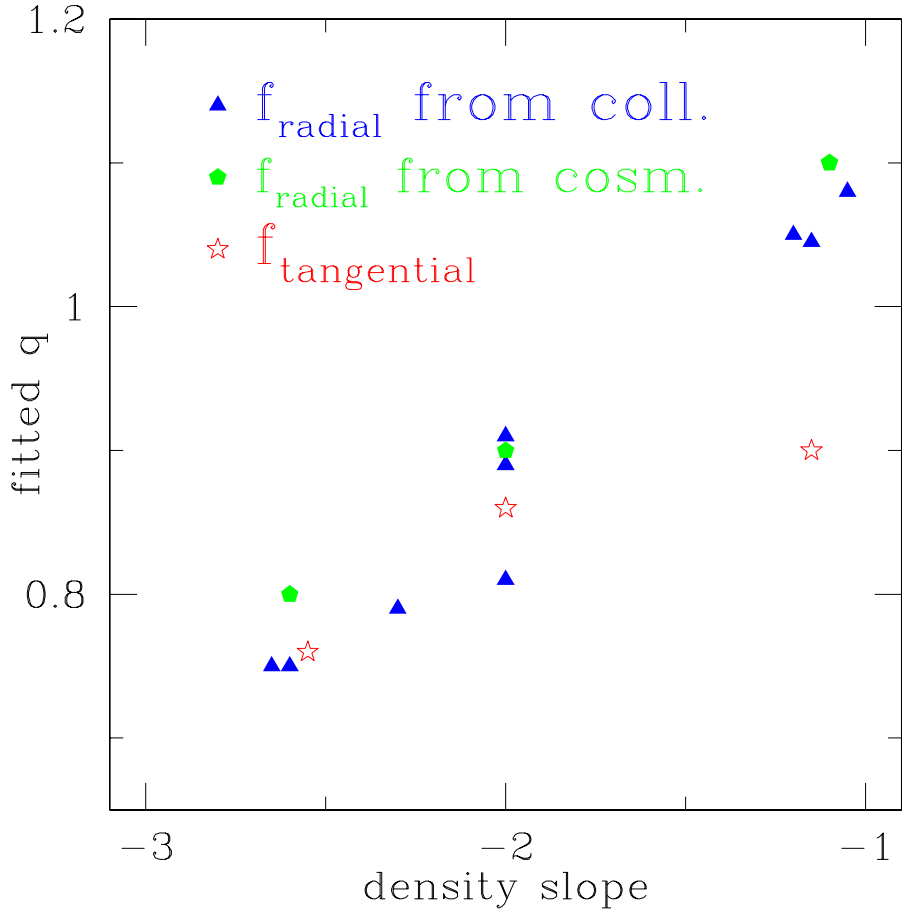}
\caption{The $q$'s which best fit the high energy tail of the radial
VDF's (blue triangles) as a function of the density slope. 
The green (filled) pentagons are fitted from the radial VDF's 
of a combination of cosmological simulations and a radial collapse
simulation. The red
(open) stars are the $q$'s which best fit the high energy tail of the
2 dimensional tangential VDF. There is a general trend, namely that
more shallow slopes are better fit with higher q. }
\label{fig:alpha.q} 
\end{center}
\end{figure}

It is clear from the figures above, that the simple shape of 
eq.~(\ref{eq:tsallis}) provides a reasonable fit to the tails 
of the radial VDF's. Furthermore, there seems to be a general
trend in the connection between density slope and the best
fitting $q$, which is shown  in figure~\ref{fig:alpha.q}
as blue (solid) triangles.

Such a trend with shallower density slope implying larger $q$'s was
suggested from simplified theoretical considerations~\cite{hansen05}.
It is interesting to note that ref.~\cite{wojtak} found that the
radial VDF of the 
inner-most bin is fit with a Gaussian $q=1$, and the intermediate
bins can be fitted with $q=0.65$, in reasonable agreement with our
findings.

\begin{figure}[hbt]
\begin{center}
\includegraphics[width=0.4\textwidth]{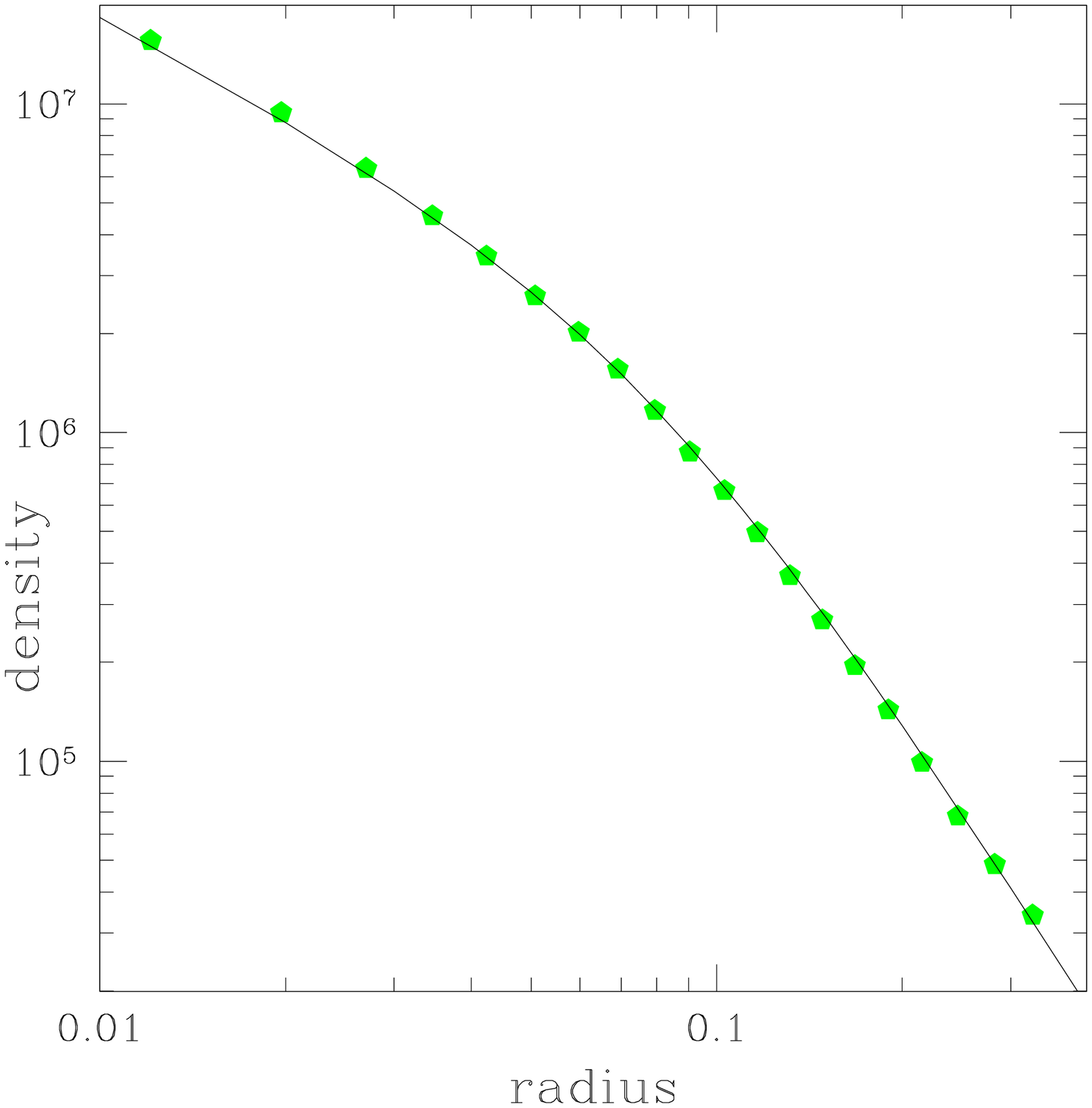}
\includegraphics[width=0.4\textwidth]{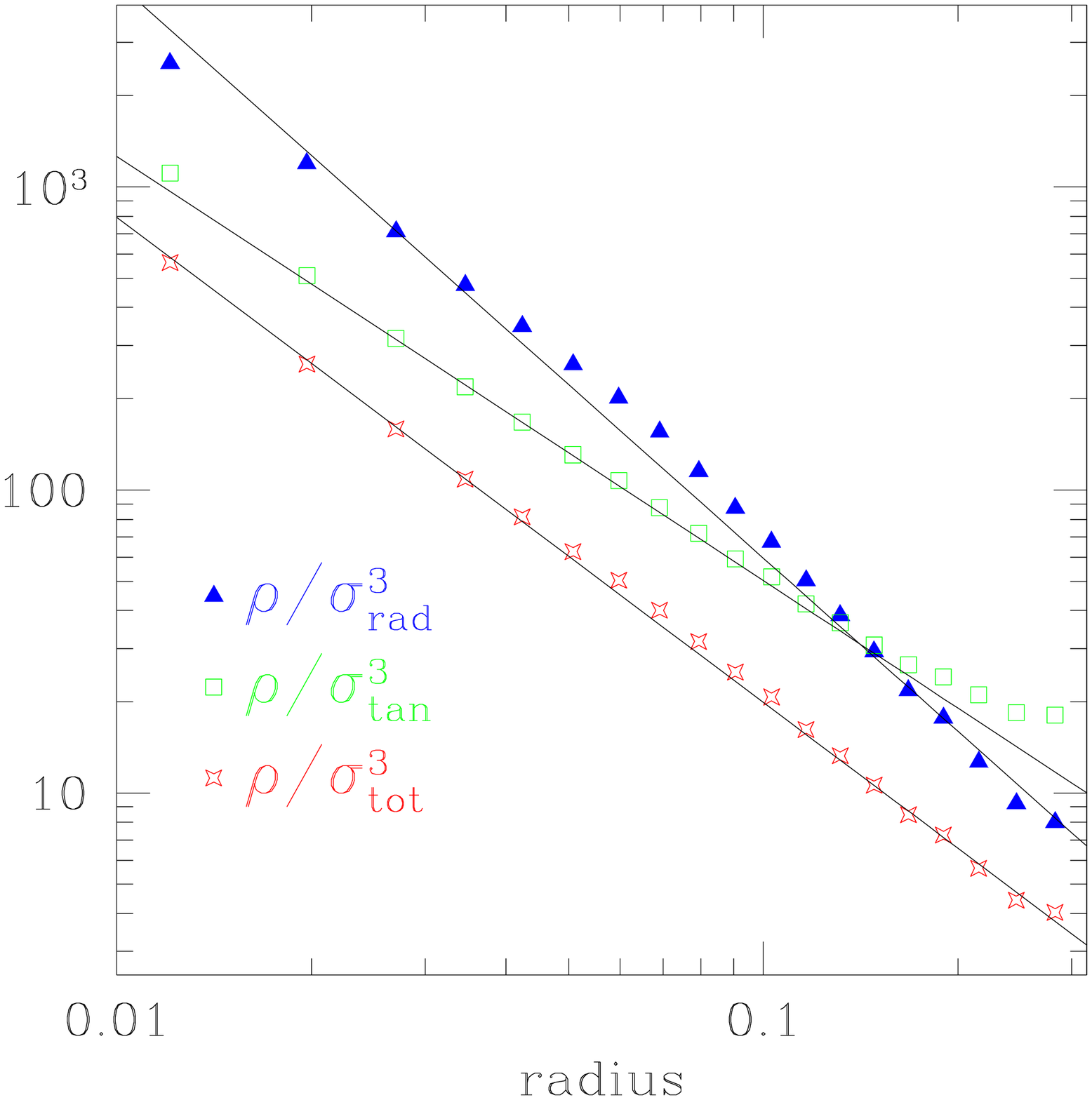}
\caption{The density and phase-space density as functions of radius for the 
radial collapse model. The black line in the left panel is of the NFW shape.
Right panel: The symbols are when using different
dispersions (radial, tangential, total), and it appears that using
the total dispersion provides a slightly better fit to a 
power-law in radius. The solid (thin) black lines are power-laws
in radius, using powers of $-1.9$ (radial), $1.4$ (tangential),
and $-1.6$ (total).}
\label{fig:phase.tophat} 
\end{center}
\end{figure}

Finally we consider simulated cosmological $\Lambda$CDM haloes that
form in a standard expanding hierarchical universe, and we analyse these
haloes in the same way as above.
We consider
two relaxed halos, each with approximately 1/2
million particles in the equilibrated region. These are the two binary
haloes taken from the Local Group simulation of \cite{calcaneo}.  One
of these haloes is strongly oblate, the other strongly prolate in density
contours.
We also perform a radial infall simulation with 1 million particles,
and analyse the resulting structure as described above. 
We first present the density and 
phase-space density of the radial infall simulation
as functions of radius (in simulation
units) in figure~\ref{fig:phase.tophat}, from which it appears
that using the total dispersion leads to a slightly better fit to
a power-law in radius, since the residuals from a power-law is
slightly s-shaped when using $\sigma_{rad}$ instead.

\begin{figure}[hbt]
\begin{center}
\includegraphics[width=0.36\textwidth]{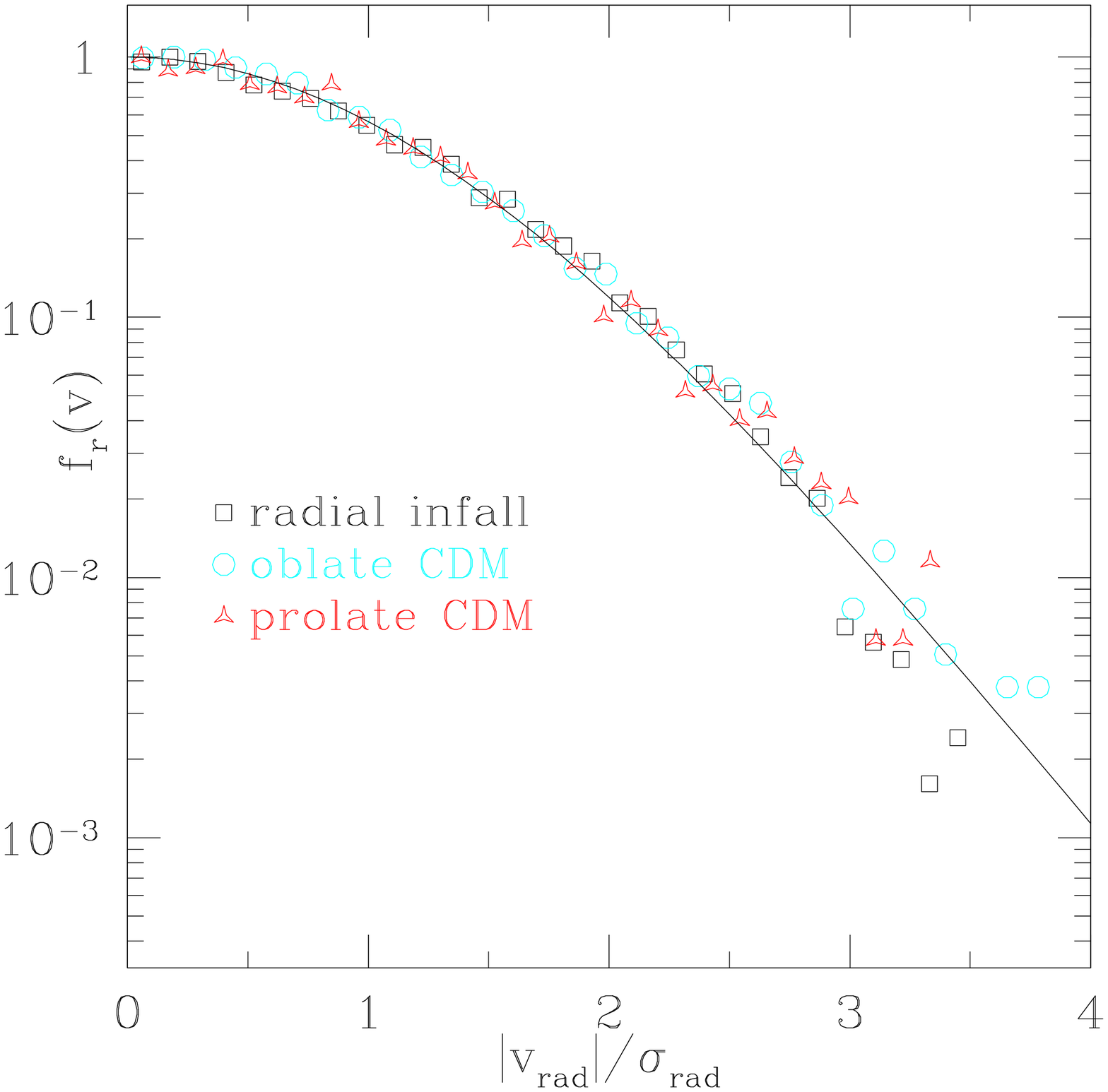}
\includegraphics[width=0.36\textwidth]{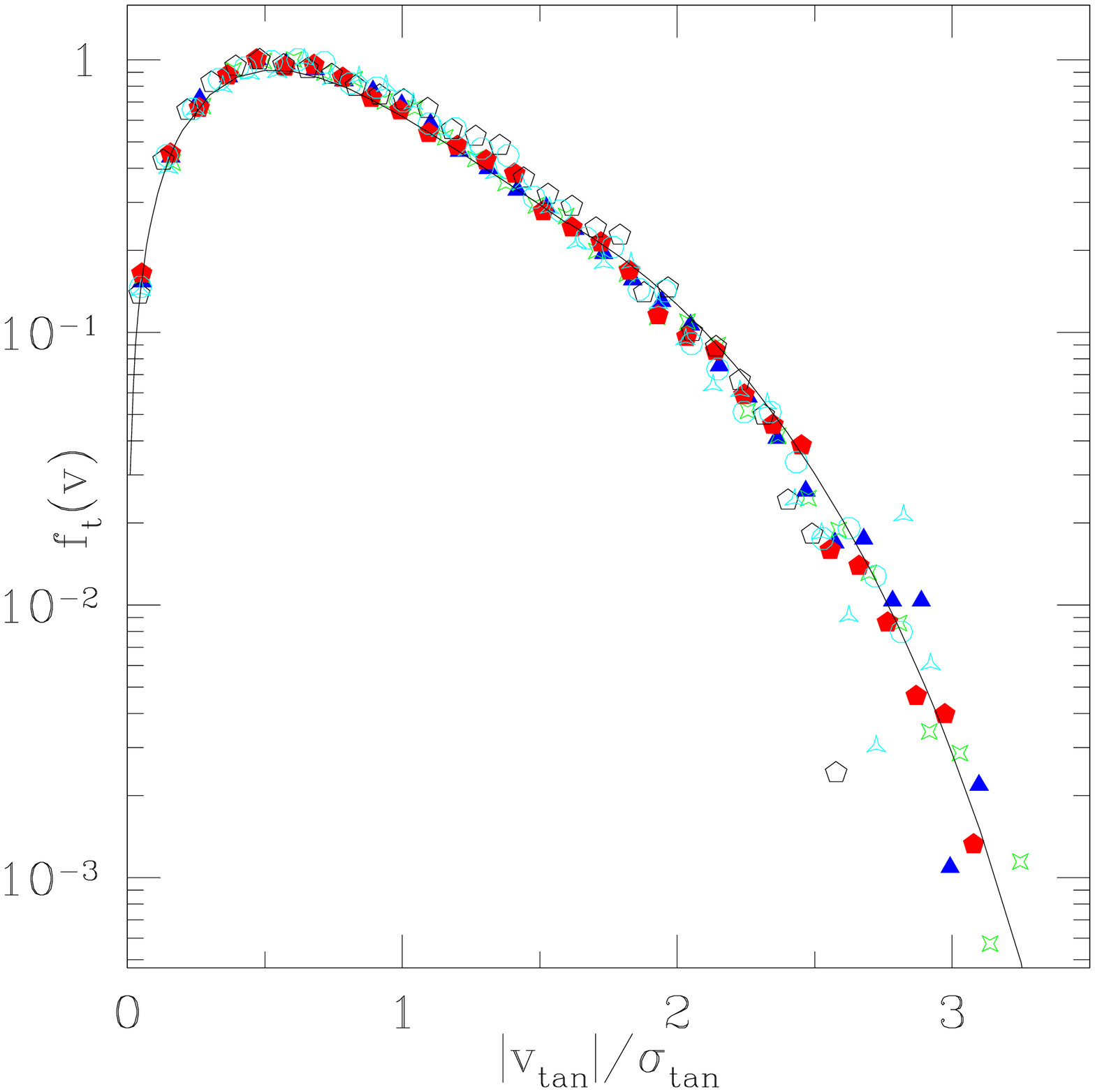}\\
\includegraphics[width=0.36\textwidth]{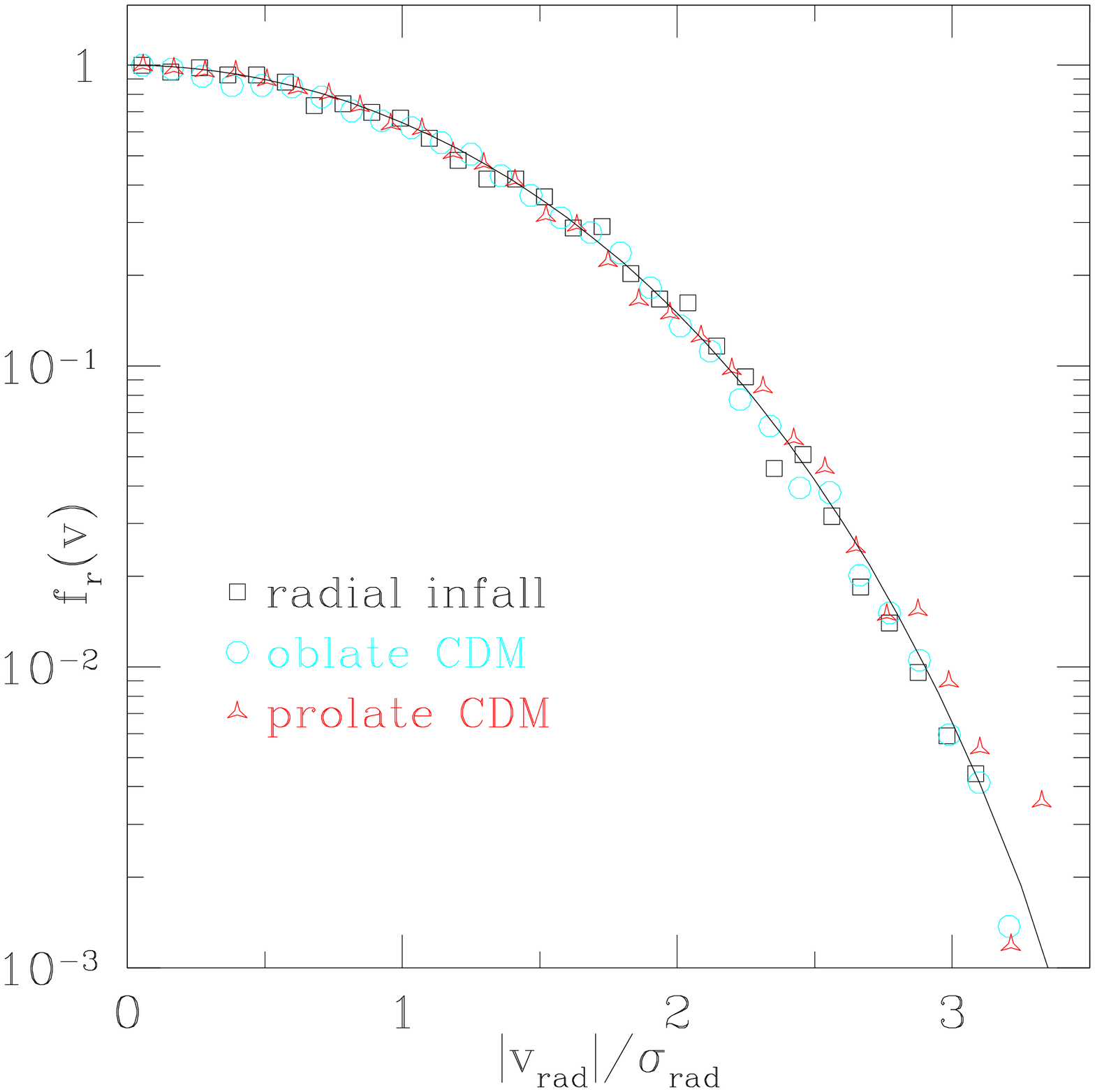}
\includegraphics[width=0.36\textwidth]{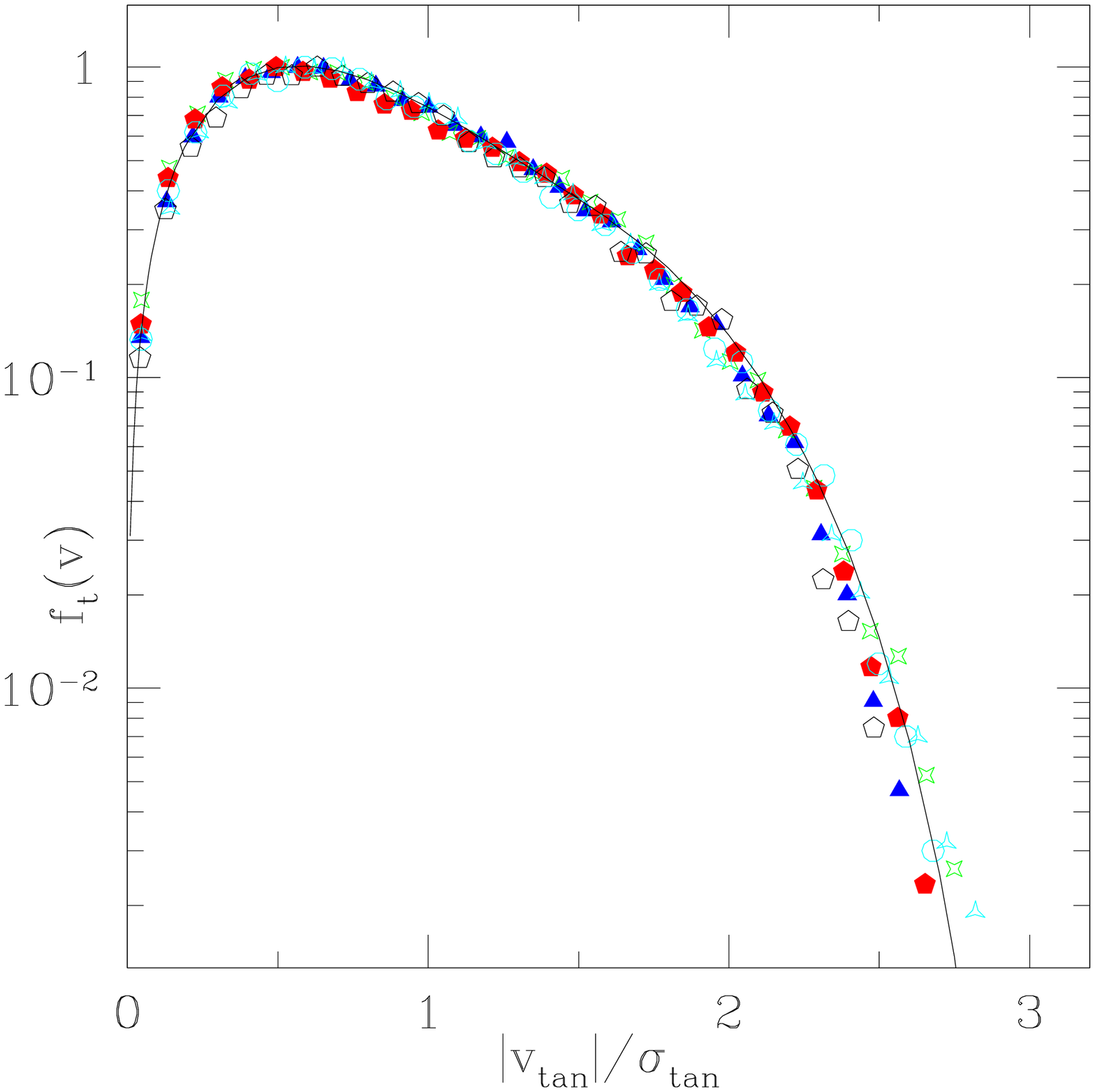}\\
\includegraphics[width=0.36\textwidth]{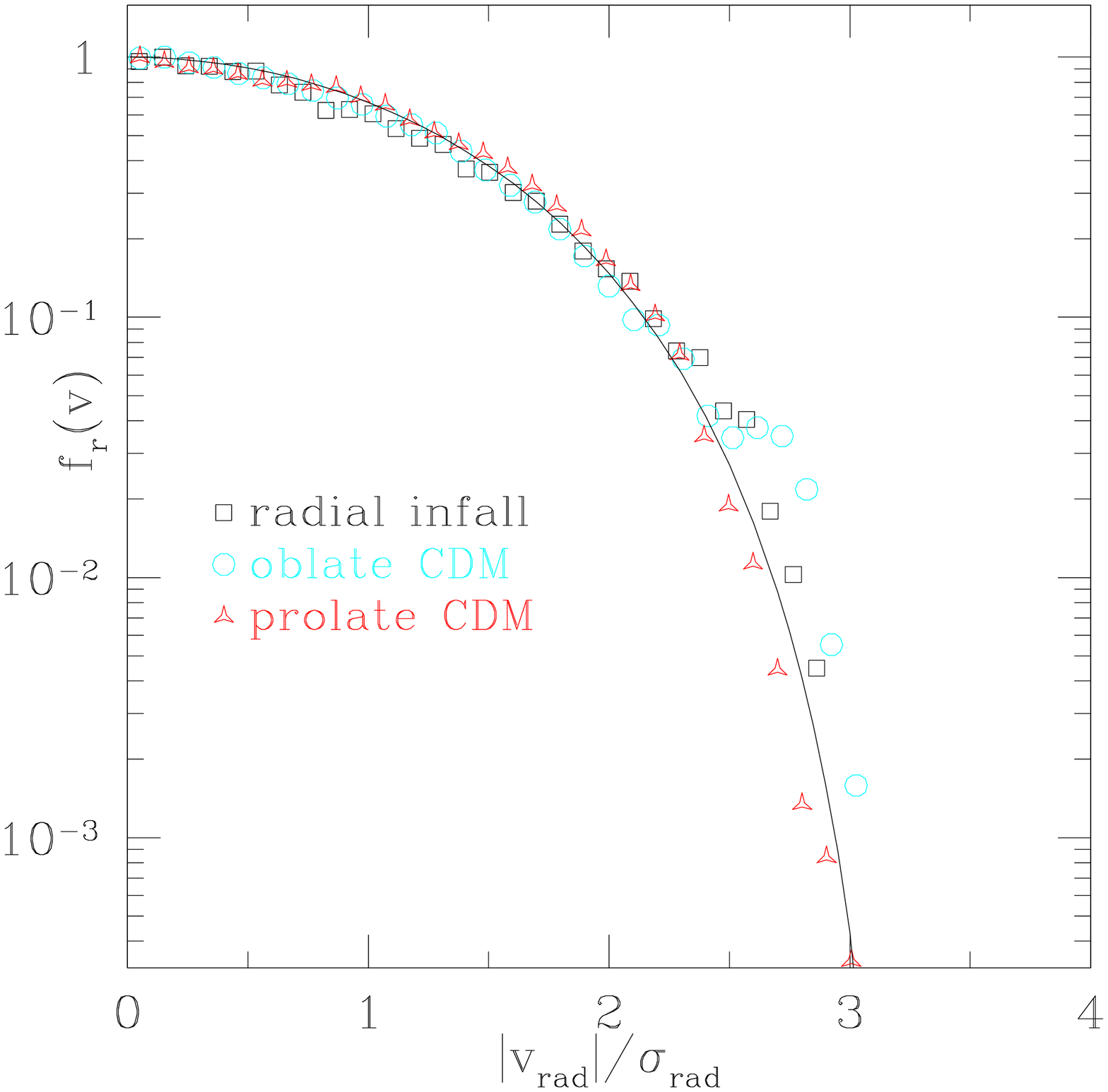}
\includegraphics[width=0.36\textwidth]{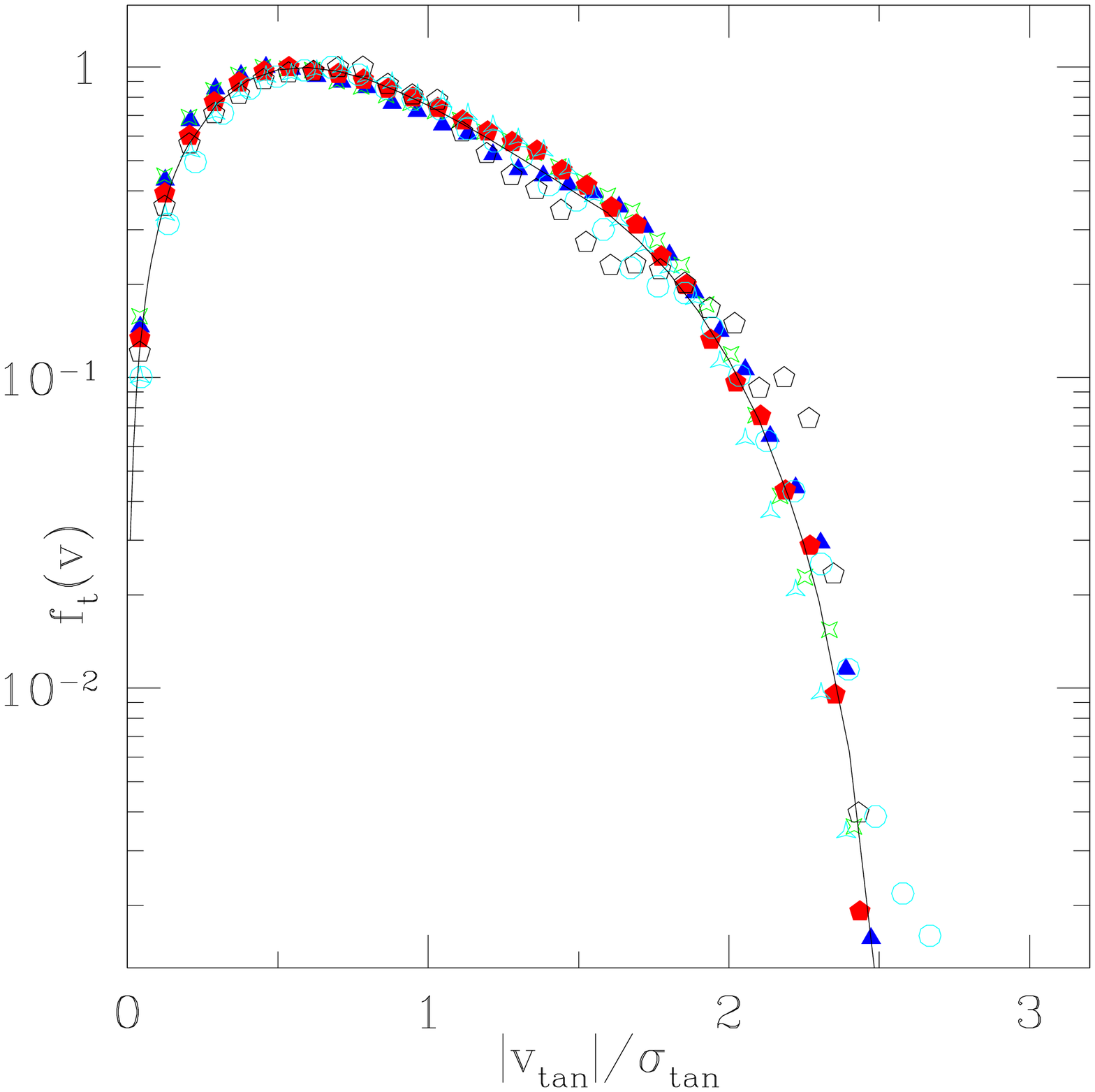}
\caption{Left column: 
The radial VDF's from a cosmological simulation (blue circles and
red triangles) and a radial infall simulation (black squares). 
Row 1: The potential bins were 
chosen near density slope of $\alpha=-1.1$, 
and the black (solid) line is of the
shape in eq.~(\ref{eq:tsallis}), using $q=1.1$.
Row 2:
$\alpha=-2$, $q=0.9$.
Row 3:
$\alpha=-2.6$, $q=0.8$.

Right column:
The tangential VDF's from all the simulations discussed
above.
The black (solid) line has two parts (separated at
$v_{\rm tan} = 1.6 \, \sigma_{\rm tan}$), each fitted with the shape
given in eq.~(\ref{eq:tsallis2}), where the low energy part uses 
$q=5/3$, and the outer region (high energy part) has a free $q$. 
Row 1: using a potential bin near the density slope of $\alpha = -1.1$,
and the black (solid) line uses $q=0.9$ in the high energy part.
Row 2: $\alpha = -2$, $q=0.86$.
Row 3: $\alpha = -2.6$, $q=0.76$.
}\label{fig:cdm} 
\end{center}
\end{figure}

For these
3 simulations we extract the VDF's in potential bins, and we 
present these 3 simulations together in the left column of
figure~\ref{fig:cdm}, 
where we are plotting the radial VDF for bins with
density slope near $-1.1, -2$ and $-2.6$ respectively. Also in these
figures do we plot black (solid) lines of the form in eq.~(\ref{eq:tsallis}),
using $q$'s of $1.1, 0.9$ and $0.8$ respectively. A potential bin
near density slope of $-2.6$ has much larger scatter since this region
is not fully mixed in the cosmological runs.
We present the corresponding 
$\alpha$'s and $q$'s as green (filled) pentagons in 
figure~\ref{fig:alpha.q}, and we find good agreement with the results
from the controlled collisions.
%


\section{Tangential VDF}

In order to discuss the shape of the tangential VDF we now 
present the tangential VDF in 3 bins from each of the simulations
described above. These are the same bins as discussed for the
radial VDF, namely one 
bin in the outer region (near $\alpha = -2.6$), 
a bin near an intermediate slope (near $\alpha = -2$),
and a bin near a shallow density slope (near
$\alpha = -1.1$).

The first impression from the right column of 
figure~\ref{fig:cdm} is that indeed all the
tangential VDF's look very similar for these different simulations:
the simulations with 4 million particles which have been run for
a very long time (red, blue and green symbols) lie very close to each
other, and the scatter in figure~\ref{fig:cdm} is from the low resolution
structures from the cosmological simulation (each with about 1/2 million
particles) and the radial infall simulation (with 1 million particles).
This scatter is most visible for the bin near a shallow profile
(top panel) where there are relatively few particles.

One should keep in mind that the initial conditions for two of these
simulations were with all the particles on completely radial orbits,
and therefore the tangential VDF's appear purely as a result of the
structure being perturbed and setteling down in a stable configuration.

When comparing figures~\ref{fig:cdm} one sees another
trend, namely that the high energy tail of the VDF becomes longer for
more shallow density profile. This was exactly what we noticed for
the radial VDF above.

These are the 2-dimensional tangential VDF's, and for a classical
gas they would have had the shape $f_t \sim v \, {\rm exp}(-v^2/v_0^2)$,
where $v_0$ is some normalizing velocity. We see clearly from the
figures that they all have a rather characteristic break, somewhere
near $v_{\rm tan} = 2 \, \sigma_{\rm tan}$, and hence a Gaussian
will give a very poor fit.

We are interested in making a phenomenological fit to the shape of
these tangential VDF, and it is natural to consider the 2-dimensional
generalization of eq.~(\ref{eq:tsallis}), namely
\be
f_t(v) = v \, \left( 1 - (1-q ) 
\, \left( \frac{v}{\kappa_2 \, \sigma_{\rm rad}} \right)^2 
\right) ^{\frac{q}{1-q}} \, ,
\label{eq:tsallis2}
\ee
however, given the break near $v_{\rm tan} = 2 \, \sigma_{\rm tan}$
this is clearly not sufficient. We will therefore fit with two 
such shapes, one in the small velocity region, and another in the
high velocity region.
It was argued in \cite{hm} that
the tangential VDF's could have the shape of eq.~(\ref{eq:tsallis2})
with $q=5/3$, and one sees immedeately that the small velocity 
region is very well fit with exactly this shape for all the bins in 
potential.
We pick the position  $v_{\rm tan} = 1.6 \, \sigma_{\rm tan}$ as
the transition velocity for all figures, and fit the high energy
region with the shape of eq.~(\ref{eq:tsallis2}), but with $q$ a
free parameter. The combined fit is shown as black (solid) lines
in the 3 figures, and these provide a very good fit to the full
VDF's. We are using $q$'s of $0.76, 0.86$ and $0.9$ 
(larger $q$ corresponds to more shallow density profile). These
results for the high energy tail of the tangential VDF's
are also shown in figure~\ref{fig:alpha.q} as red (open) stars.
Again there is a general trend showing that more shallow density
profiles give tangential VDF's which have power-law tails given
by larger $q$'s.

\section{Full VDF}

Having identified that both the radial and the tangential
distribution functions have the same universal shapes for
different simulations, for potential bins which are at the
same value of the density slope, we can now also plot the
full (3 dimensional) VDF. 
Also cosmological simulations have considered
the full VDF and found that a Gaussian in general does not give a good 
fit to the VDF~\cite{calcaneo,diemand}.

A comment on the uniqueness of the connection between the VDF's and
the density is in order at this point. The density is simply given as
an integral over the full VDF, $\rho = \int f \, d^3v$, however, this
is completely independent of the anisotropy. If the anisotropy is
zero, then this equation can be inverted for spherical systems (by the
Eddington method) to give the VDF as a function of the density
profile. This inversion gives a unique VDF. However, for an
anisotropic and non-spherical system this inversion is almost
certainly not unique, and it is therefore not obvious that the VDF
should be universal between these different simulations.  On a more
philosophical note, one should keep in mind that the individual
particles are gravitationally scattered off each other while the
structures are forming: the individual particles only feel the local
particles and the local change of density and potential.  It is
therefore the VDF's which are truely universal. The integrated
quantities, namely density and anisotropy, are universal because of
the universality of the VDF's.  If we are ever going to truely
understand the universality of density and anistropy, then we should
take a step back and understand the universality of the VDF's.

\begin{figure}[hbt]
\begin{center}
\includegraphics[width=0.8\textwidth]{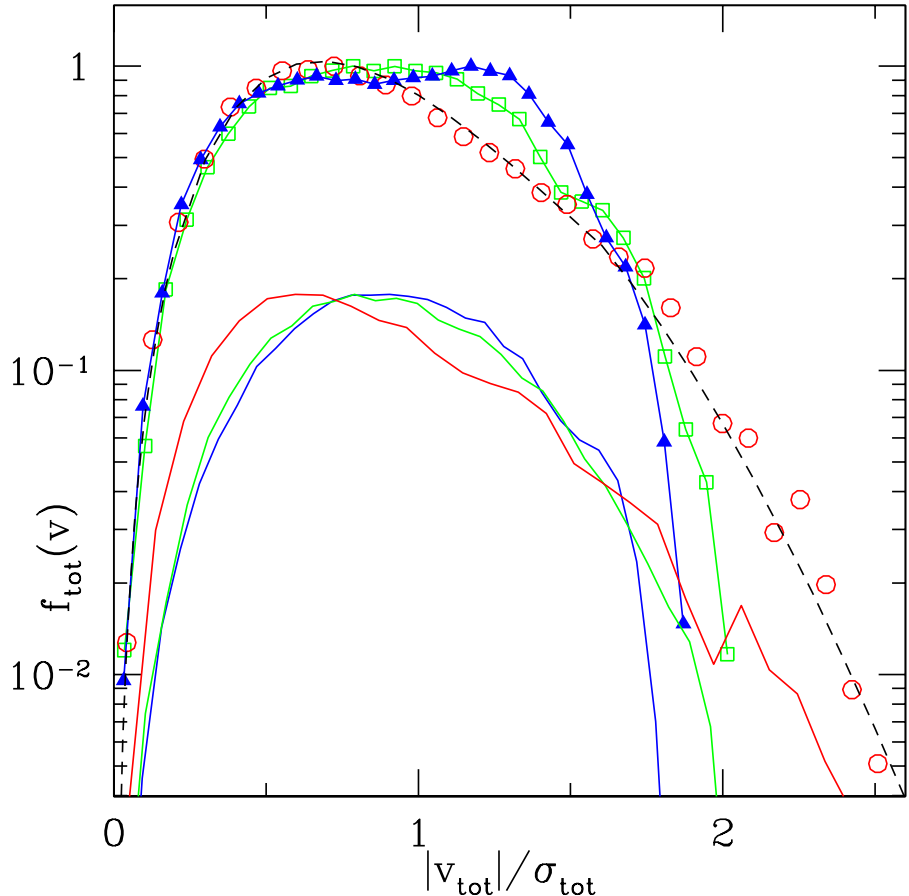}
\caption{The total 3-dimensional VDF for 3 potential bins in the simulation
of a repeated headon collision between initially isotropic NFW
structures. The potential bins are chosen near density slope of
$\alpha \approx -1.1$ (open red circles), 
$\alpha \approx -2$ (open green squares)
and $\alpha \approx -2.6$ (filled blue triangles).
The solid lines which are shifted downwards for visibility are from 
the isotropic NFW structure, and do not exhibit the double-hump.}
\label{fig:full}
\end{center}
\end{figure}

Instead of plotting for all the
simulations, we will only show the full VDF for one, namely the repeated
head-on collision between initially isotropic NFW structures.
This is shown in figure~\ref{fig:full}, where
the potential bins are chosen near density slope of
$\alpha \approx -1.1$ (open red circles), 
$\alpha \approx -2$ (open green squares)
and $\alpha \approx -2.6$ (filled blue triangles).
%

In the spatially central region with shallow density profile there is zero
anisotropy, and the full VDF can again be reasonably fitted by
two functions of the shape $f_{\rm tot} = v^2 \, f_r$. This is
shown with the dashed (black) line going through the red circles
(using different $q$'s in the high and low energy region, $q_1 \approx 1.3,
q_2 \approx 1$).
In the spatial regions with non-zero anisotropy the full VDF looks much more
complicated, in particular, in regions with large anisotropy is
develops a double-hump shape (see blue filled triangles in 
figure~\ref{fig:full}).  For comparison we show the corresponding full
VDF from the initial conditions of a stable NFW structure with zero
anisotropy as solid lines shifted downwards for visibility
(regions corresponding to the same density slope as for the 
repeated head-on collision). These
initial conditions clearly do not have the double-hump, since they
are isotropic.

\section{Conclusions}

We have identified a universality of the velocity distribution
function (VDF) of equilibrated structures of collisionless,
selfgravitating particles through a set of controlled numerical
experiments.  We find that the radial and tangential VDF are universal
in the sense that they depend only on the dispersion (radial or
tangential) and the local slope of the density.  This universality of the
VDF is non-trivial, since the final structures are not spherical
and isotropic.

Our simulations include controlled collision experiments between
structures which are initially isotropic as well as highly
anisotropic, radial infall collapse, and structures formed in
cosmological simulations.

The high energy tail of the radial VDF is well approximated by
the shape given in eq.~(\ref{eq:tsallis}), however, for regions
with non-zero anisotropy (steep density slopes) there are corrections
in the low-energy part of the VDF.

The tangential VDF is well described by a combination of two
functions of the shape of eq.~(\ref{eq:tsallis2}), where the
low-energy part has $q=5/3$, and the high-energy part has a $q$
similar to the one describing the radial VDF.

Whereas the general trend, namely that steeper density slope 
implies small $q$, is in agreement with theoretical expectations,
then the actual shape of the VDF's is not understood yet. Our
numerical results may hopefully inspire further theoretical
understanding of the VDF of collisionless particles.

\ack 
We sincerely thank J{\"u}rg Diemand for pointing out an important
mistake in an early draft of this paper.  It is a pleasure to thank
Victor Debattista and David Merritt for interesting suggestions, and
Ewa {\L}okas and Gary Mamon for useful comments on an early draft.  We
thank Stelios Kazantzidis for the use of MakeHalo.  SHH thanks the
Tomalla foundation for support.

\label{lastpage}

\end{document}